\documentclass[reprint, amsmath, amssymb, aps,pre]{revtex4-2}

\usepackage{graphicx}
\usepackage{tikz}
\usepackage{dcolumn}
\usepackage{array}
\usepackage{todonotes}
\usepackage{bm}
\usepackage{siunitx}

\begin{document}

\title{A Vibrated Compacting Granular System: A DEM Light Scattering Comparison}

\author{Linnea Heitmeier }
\email{linnea.heitmeier@dlr.de}

\author{Jan Philipp Gabriel}
 \email{jan.gabriel@dlr.de}
\affiliation{
German Aerospace Center (DLR) Institute of Frontier Materials on Earth and in Space,
 51170 Cologne, Germany
}

\date{\today}

\begin{abstract}
We perform Discrete Element Method (DEM) simulations of granular particles (polystyrene spheres) vibrated inside a cubic container. The study investigates the evolution of the packing fraction with and without rotational friction at different shaking amplitudes. The mean-squared displacement (MSD) is used to analyze the particles' diffusive, subdiffusive, and superdiffusive behavior. By monitoring both the dynamics and density evolution, one can observe the system's glassification. The comparison with experiments shows that the MSDs from the simulations are significantly higher than the MSDs measured by Diffusing Wave Spectroscopy (DWS) \cite{kunzner2025dynamics}. Following our finding that the rotational MSD is of the same order of magnitude as the MSD measured in DWS experiments, we propose that the experimental signal is not dominated by translational motion but rather by rotational particle dynamics. This provides access to a relevant particle property that has previously been difficult to measure directly. Finally, we conclude that the system reaches a dynamically constrained state well before random close packing, with particle displacements already below the Lindemann length.
\end{abstract}

\maketitle

\section{Introduction}

Granular matter is ubiquitous in nature and industry; examples include snow, nuts, coal, and rice \cite{jaeger1996physics, duran2012sands, dufresne2012granular, rosato1987brazil}. It also plays a central role in applications such as powder-based additive manufacturing, where the flow properties of the filament, a granular matter, are of high importance for the resulting printing quality \cite{sweeney2017characterizing, roy2024role, asaf2024granular}. Earlier studies have shown that dense granular matter exhibits behavior comparable to that of dense liquids and glass-forming systems \cite{liu1998jamming, zhang2005jamming,berthier2009glasses,siemens2010jamming, zhang2010jamming, behringer2018physics}. Similar to glassy systems, even small changes in packing fraction can lead to significant changes in the dynamics of the system. In contrast to molecular and colloidal fluids, where thermal fluctuations provide a natural source of energy, granular systems are athermal. Hence, energy must be supplied externally. A common method for inducing granular dynamics is mechanical vibration. Vibrated granular systems have been extensively investigated experimentally \cite{kunzner2025dynamics, mayo2025observing, kollmer2020migrating, Scalliet2015Cages, eshuis2007phase, kudrolli2004size, nicolas2000compaction, knight1995density, nowak1998density, makse2000packing, gerlach1927messung}. Although several simulation studies of such systems have been reported \cite{el2021theories, hadi2024modelling, di2021coarse, hashemnia2018study}, they do not permit a direct comparison with our experiments. In particular, these studies did not simultaneously quantify the compaction, translational mean-squared displacement (MSD), and rotational MSD.

In this work, we address this gap by performing Discrete Element Method (DEM) simulations of a vibrated granular system inspired by a light-scattering experiment using Diffusing Wave Spectroscopy (DWS) on 140 µm polystyrene spheres shaken at 100~Hz inside a cubic container (4~mm$^3$) \cite{kunzner2025dynamics}. A puzzling observation in the experiment is that the extracted mean squared displacements (MSDs) suggest caging lengths of only 0.07\% of the particle diameter \cite{kunzner2025dynamics}. Similar results are found in densification experiments on 140 µm polystyrene spheres under low gravity on board the ISS \cite{mayo2025observing}. The densification process is monitored by the MSD extracted by DWS and implies earlier glassification and jamming under low gravity. The low caging lengths are not in line with Mode-Coupling Theory (MCT) predicted caging at approximately 10\% of the particle diameter \cite{sperl2005nearly,sperl2012single} following the melting concept related to the Lindemann length \cite{Lindemann1910}. This opens the question, also relevant to other DWS experiments \cite{menon1997diffusing, menon1997particle, biggs2008granular, hanotin2013dynamics, hanotin2015viscoelasticity, kim2002solid,kim2005jamming, zivkovic2011scaling, M12}, if the DWS Method is truly measuring the translational dynamics in these cases. 

\begin{figure}[t!]
\centering
\begin{tikzpicture}[>=latex, thick]
\node[inner sep=0pt, minimum width=5cm, minimum height=3cm] (img) 
{\includegraphics[width=5cm]{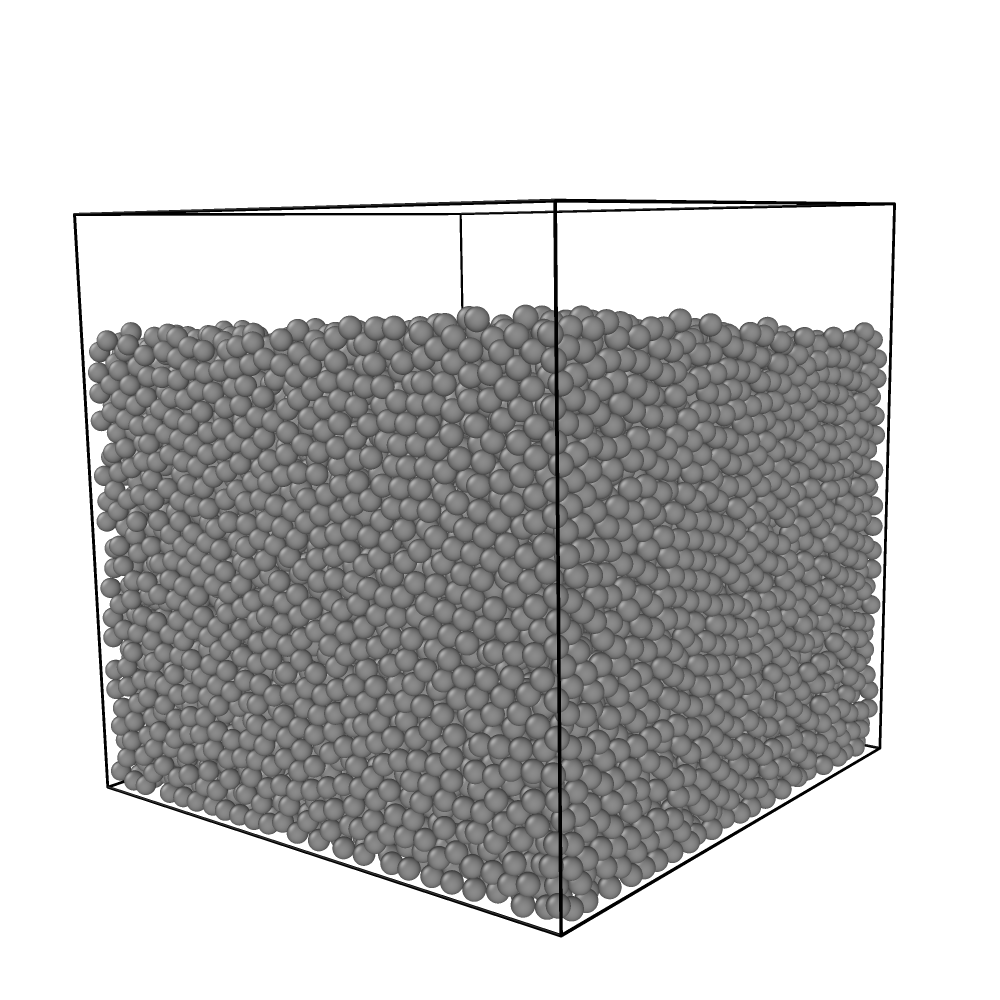}}; 
\draw[->, very thick] ([xshift=-0.5cm]img.west) -- ++(0,-1.2cm)
    node[below] {$F_g$};
\draw[<->, very thick] ([xshift=0.5cm]img.east) ++(0,-0.6cm) -- ++(0,1.2cm)
    node[midway, right] {$(A, \omega)$};
\end{tikzpicture}
\caption{Schematic representation of the simulation setup after sedimentation. The gravitational force $F_g$ acts downward, while the container walls oscillate vertically with amplitude $A$ and angular frequency $\omega$.}
\label{fig:setup}
\end{figure}

To resolve this discrepancy, we monitor the packing fraction as well as the translational and rotational MSDs during the shaking process. While translational MSDs are commonly studied, rotational MSDs are less frequently reported; examples can be found in Refs. \cite{elizondo2020arrested, Scalliet2015Cages, gerlach1927messung} and references therein. We discuss the role of rotational friction in determining the rotational MSD and address the challenges associated with densifying the system when rotational friction is included in the simulations. In the experiment, the particles densify up to packing fractions of 62\% \cite{kunzner2025dynamics}, consistent with reported values for random close packing in systems with friction \cite{silbert2001granular, jerkins2008onset, liu2017equation}. We analyze the diffusive, subdiffusive, and superdiffusive regimes observed in the calculated MSDs and discuss their relation to the MSDs extracted from DWS measurements \cite{kunzner2025dynamics} and there experimental implications for other DWS experiments like the densification experiment under low gravity on board the ISS \cite{mayo2025observing}.

\section{Method}

Discrete Element Method (DEM) simulations were performed using the open-source molecular dynamics software \texttt{LAMMPS} \cite{LAMMPS}. The interactions between particles were modeled using a Hookean contact law, where particles experience a force only when they overlap. The corresponding pairwise force between particles $i$ and $j$ is given by \cite{brilliantov1996model,silbert2001granular,zhang2005jamming}:
\begin{align}
    \mathbf{F}_\text{ij, Hooke} = 
    (k_n \, \delta \, \mathbf{n}_{ij} - m \, \gamma_n \, \mathbf{v}_n)
    - (k_t \, \Delta \mathbf{s}_t + m \, \gamma_t \, \mathbf{v}_t),
\end{align}
where $\delta$ is the overlap between two particles, $k_n$ and $k_t$ denote the elastic constants in the normal and tangential directions, respectively, and $\gamma_n$ and $\gamma_t$ are the corresponding damping coefficients. The variable $m$ represents the effective particle mass defined by $m= \frac{m_i m_j}{m_i+m_j}$, while $\mathbf{v}_n$ and $\mathbf{v}_t$ are the normal and tangential components of the relative velocity. The vector $\Delta \mathbf{s}_t=\int dt \space v_t$ accounts for the tangential displacement between two particles and includes the history of tangential contacts. We applied the static yield criterion $x_\mu$, which limits the tangential force by $F_t \leq x_\mu F_n$. Here, $F_n$ is the normal force of the interaction. If not mentioned otherwise, it is set to 0. The case of $x_\mu=0$ is called the frictionless case, while the case of $x_\mu\neq 0$ is called the frictional case. 

Initially, $N$ monodisperse particles were placed randomly within a rectangular box of dimensions $L_x \times L_y \times L_z$. A gravitational acceleration of magnitude $g$ was applied in the negative $z$-direction to induce particle settling and segregation. 
The boundaries of the simulation box were modeled as walls interacting with the particles through the same Hookean potential, but with different interaction parameters (see Table~\ref{tab:parameters}). The setup after the sedimentation is illustrated in Figure \ref{fig:setup}. 

Subsequently, a sinusoidal excitation was imposed by oscillating all walls with amplitude $A$ and angular frequency $\omega=2\pi f$, such that the $z$-position of the walls $w_z$ at time $t$ was given by 
\begin{align}
   w_z(t) =  w_z(0)+ A \cdot \sin\left(\omega t \right)
\end{align}
Similar to \cite{kunzner2025dynamics}, we make use of the non-dimensional gravitational acceleration 
\begin{align}
    \Gamma=\frac{A \omega^2}{g}
\end{align}  
During the shaking process, we monitored the evolution of the packing fraction and the \textit{mean-squared displacement} (MSD). For the system with non-vanishing tangential forces, we also tracked the \textit{rotational mean-squared displacement} (rMSD).

The packing fraction $\phi$ was determined as the ratio of the volume occupied by the particles and the ''effective" volume, defined by the minimal and maximal particle coordinates in each direction: 
\begin{align}
    \phi = \frac{V_\text{particles}}{V_\text{total}}
\end{align}
We checked that the results we obtain from this method are consistent with an alternative approach, which calculates the density of the system in dependence on the $z$-coordinate. 

The translational mean-squared displacement was calculated as
\begin{align}
    \text{MSD}(t+\Delta t) = \langle \, |\mathbf{r}(t+\Delta t) - \mathbf{r}(t)|^2 \, \rangle,
    \label{eq:msd}
\end{align}
where $\mathbf{r}(t)$ denotes the particle position at time $t$, and the brackets $\langle \cdot \rangle$ indicate an ensemble average over all particles. For the calculation of this quantity, we used the built-in methods of \texttt{LAMMPS}, which calculates the MSD \cite{lammpsmsd}. \\
The rotational mean-squared displacement was obtained from the time-integrated angular velocity $\boldsymbol{\omega}(t)$ of each particle, scaled by the particle diameter $d$:
\begin{align}
    \text{rMSD}(t) = \Big\langle \left( \frac{d}{2} \,  \int_0^t \boldsymbol{\omega}(t') \, dt' \right)^2\Big\rangle.
    \label{eq:rmsd}
\end{align}
This quantity provides a measure of the particle’s reorientation $\alpha$ due to rotational motion. 
Note that the prefactor $d/2$ converts the angular displacement into the corresponding arc length travelled by a point at the particle surface. The rMSD does not saturate at a specific value because it is based on the time-integrated magnitude of the angular velocity, which accumulates the total rotational displacement. In contrast, a quantity based only on the particle orientation angle would be periodic and would saturate after a full rotation at $2\pi$.
Note that the shapes of formula \eqref{eq:msd} and \eqref{eq:rmsd} differ from each other. The reason for this is that with our simulations, we cannot directly access the particles' orientations, but only their angular velocities. By integrating the angular velocities, we get the rMSD, which is in terms of the orientation given by $\text{rMSD}(t)= \frac{d}{2} \, \langle \alpha(t+\Delta t) -\alpha(t) \rangle $. Note that with this definition, the rMSD can take values larger than $2 \pi$. Consequently, it does not trivially saturate at one specific angle.

The simulation parameters are summarized in Table~\ref{tab:parameters}. The material properties were chosen to resemble those of polystyrene particles (Dynoseeds 140), adopting parameter values consistent with previous studies \cite{d2021gravity,mair2007nature,vescovi2016merging,malone2008determination}. Note that despite decades of research, the exact parameters that should be used in the Hookean spring model are still unknown \cite{di2004comparison}. All results are averaged over 7 different initial configurations.

\newcolumntype{C}{>{$}c<{$}} 
\begin{table}
\centering

\begin{tabular}{|c|C|}
     \hline
     quantity & \text{value}  \\ \hline  \hline
     particle diameter $d$ [m]& 140\times 10^{-6} \\
     density $\rho$ [kg/m$^3$] & 1600 \\
     number of particles $N$ [-] & 20000 \\
     box lengths $L_x$, $L_y$, $L_z$ [m] &
     0.004, 0.004, 0.004\\
     gravitational acceleration $g$ [m/s$^2$] & 9.81 \\
     elastic stiffness $k_n$, $k_t$ [N/m] & 33.81,\ 4.52 \\
     damping constant $\gamma_n, \gamma_t$ [kg/s] & 1.01\times10^{4},\ 9.08\times10^{3} \\
     wall stiffness $k_n^{\text{wall}}$, $k_t^{\text{wall}}$ [N/m]
     & 90.55,\ 1.22\times10^{4} \\
     wall damping  $\gamma_n^{\text{wall}}, \gamma_t^{\text{wall}}$ [1/s] &
     1.97\times10^{4},\ 1.77\times10^{4} \\
     shaking frequency $f$ [Hz]& 100 \\
     shaking amplitude $A$ [m] & 1\times10^{-5} \text{ to } 3\times10^{-5}\\
     dynamic yield criterion [-] & 0; 0.1 
     \\ \hline
\end{tabular}

\caption{Parameters used in the simulations. With the given units, the values are comparable with those of polystyrene \cite{di2004comparison}.}
\label{tab:parameters}
\end{table}

A direct comparison with experiments can be made using scattering techniques that measure density fluctuations, such as photon correlation spectroscopy \cite{berne2000dynamic}. Since the scattering cross-section of large particles is very small for visible light, experiments often operate in the multiple-scattering regime. In this case, the diffusing wave spectroscopy (DWS) approximation allows one to relate the electric field autocorrelation function $g_1(t)$ to the mean-squared displacement $\langle r^2 \rangle$ \cite{Brown}:
\begin{equation}
    \label{eq:FieldAutoCorr}
    g_1(t) \propto \exp\left(-C \, \langle r^2(t) \rangle \right),
\end{equation}
where $C$ is a constant that depends on the scattering geometry and optical properties of the system\cite{Brown,mayo2025observing,kunzner2025dynamics,M12}.

If the scattering centers are not the particle centers themselves but internal density inhomogeneities, such as air bubbles within polystyrene particles and surface roughness, the measured signal may predominantly reflect rotational rather than translational motion. In this case, the accumulated phase shifts \cite{M12} lead to an effective relation where the translational mean-squared displacement is replaced by the mean-squared angular displacement. We approximate the effective displacement of scattering centers within a particle of radius $R$ as
\begin{equation}
    \label{eq:esti}
    \langle \Delta r^2 \rangle  \approx R^2 \, \langle \Delta \theta^2 \rangle 
\end{equation}
assuming that the MSD decay is dominated by scattering events originating near the particle surface, even when the scattering centers are uniformly distributed throughout the particle, since these events contribute most strongly to the decay.

Consequently, rotational motion produces an effective displacement signal that is interpreted as an rMSD. This interpretation is consistent with experimental observations for optically opaque scattering particles such as polystyrene \cite{mayo2025observing,kunzner2025dynamics}, but does not apply to fully transparent or purely reflecting particles, where scattering occurs predominantly at the particle surface. We will find in the simulation that the ballistic part of the MSD is approximately $R^2$ times the rMSD.

\section{Results}

\begin{figure}[t!]
\centering
\includegraphics[width=\linewidth]{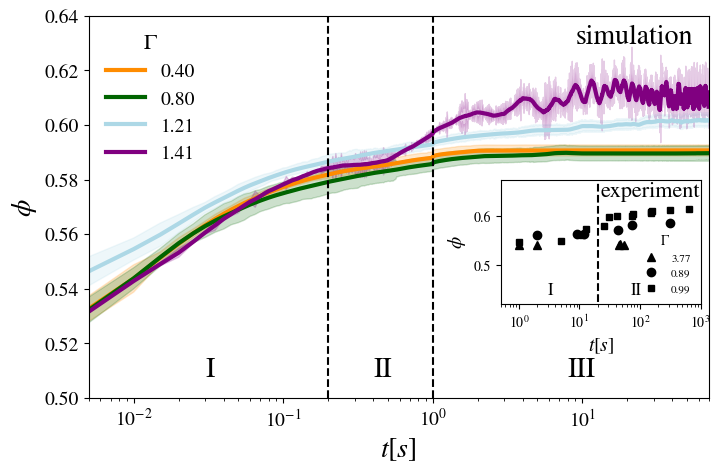}
\caption{Density evolution of the system for  $A=3.5 \times 10^{-5}$ (violet), $A=3 \times 10^{-5}$ (blue), $A=2\times 10^{-5}$ (green), and $A=1\times 10^{-5}$ (orange). All lines correspond to the average over at least 7 configurations. The light areas correspond to the uncertainty. In the inset, the black data points correspond to the experimental data from \cite{kunzner2025dynamics}. }
\label{fig:density}
\end{figure}

During the shaking process, we tracked the packing fraction of the particles shown in Figure \ref{fig:density}. The densification strongly depends on the shaking amplitude. For very small amplitudes, the system densifies very slowly; for intermediate amplitudes, it densifies more quickly; and for large amplitudes, no densification is observed. This behavior can be explained as follows: Shaking the system injects energy into the system. If a large amount of energy is supplied, the particles become fluidized; if the energy input is smaller, the particles can use the energy to rearrange and approach a more favorable configuration. 

The densification appears to be divided into three regimes: (I) fast densification, (II) slower densification, and (III) a regime in which the system does not densify and is prevented from exploring the full energy landscape. The frictionless simulations exhibit significantly faster densification, occurring on the order of seconds, whereas the experimental system densifies on a timescale of approximately 100 seconds when approaching the transition between regimes I and II. Nevertheless, this transition is clearly observed in both the simulations and the experiments. In regime III, the experiment continues densifying up to the measured timescale of 10\,000 seconds, whereas simulations that use the same number of particles as the experiments are currently computationally infeasible.

\begin{figure}[t!]
\centering
\includegraphics[width=\linewidth]{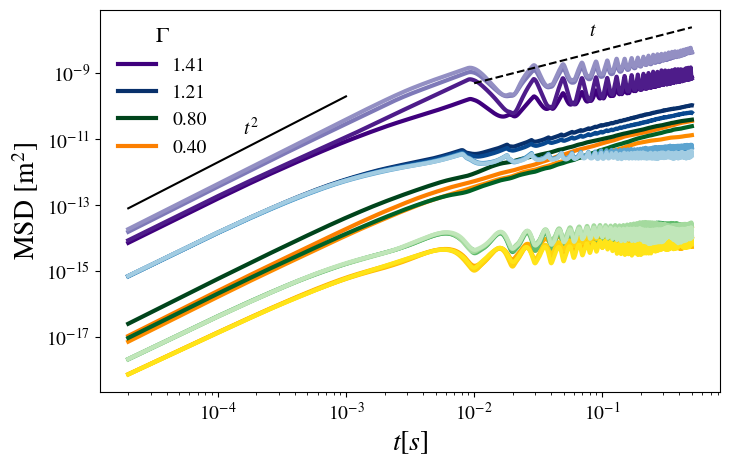}
\caption{MSDs of the particles during the shaking process. Bright lines correspond to earlier times, while dark ones correspond to late times. We show the MSDs for different shaking amplitudes $A=3 \times 10^{-5}$ (blue), $A=2\times 10^{-5}$ (green), and $A=1\times 10^{-5}$ (orange). As indicated in the legend, the gray lines for orientation correspond to $t$, and $t^2$, respectively. 
}
\label{fig:MSD}
\end{figure}

During the shaking process, the mean squared displacement (MSD, see Figure \ref{fig:MSD}) exhibits the characteristic behavior of glassy systems. Two temporal regimes are observed: at short times, the MSD increases quadratically, indicative of ballistic motion, while at long times it grows linearly, reflecting diffusive behavior. The superimposed oscillations, which can be seen in the long-time behavior, are a signature of the shaking movement. As the density increases over time, the sub-diffusive MSD evolves, and a plateau emerges, corresponding to particle caging. This plateau becomes more pronounced at higher densities, reflecting a slowdown of the dynamics and a reduction in particle mobility. 

\begin{figure}[t!]
\centering
\includegraphics[width=\linewidth]{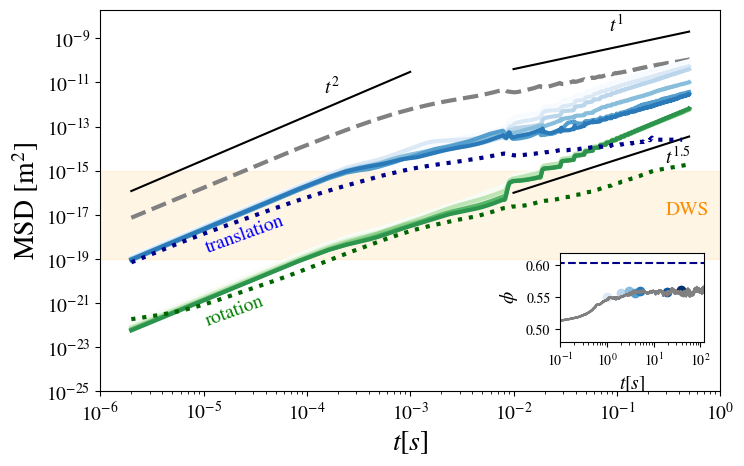}
\caption{Comparison of translational MSD and rotational MSD for the case of $x_\mu=0.1$ and $\Gamma=1.21$, i.e., tangential forces play a role here. The dashed line is the case $x_\mu=0$ for a comparison. Solid lines correspond to the case of $x_\mu=0.1$. The dotted lines correspond to simulations with $x_\mu=0.1$, but a higher packing fraction ($\phi = 0.604$). For this, we took a configuration from the case $x_\mu$, as discussed in the text, and later switched on the effect of friction. The inset shows the evolution of the density profile, with scatter points at the times where (r)MSDs are measured. The horizontal line corresponds to $\phi = 0.604$. The yellow-shaded area corresponds to the lengthscales which can be measured using DWS.} 
\label{fig:rotmsd}
\end{figure}

Rotational friction is necessary to calculate the rMSD. Figure \ref{fig:rotmsd} shows the corresponding results for different frictions. As additional friction increases the computational demands, we are not able to study the rMSD over the entire densification process. Nevertheless, we can use the densification reached without rotational friction to reach densities relevant to the experiment and calculate MSDs there. In the plot, solid lines correspond to the MSDs (and rMSDs) measured during the simulation, dotted lines correspond to the case where we used a dense system as a starting configuration ($\phi = 0.604$). The behavior of the rMSD strongly depends on the parameters $k_t$ and $\gamma_t$, which are largely unknown and might be calibrated with rotational information obtained by DWS.

Notably, a comparison with experiments \cite{kunzner2025dynamics} shows that the order of magnitude of the MSD obtained experimentally differs by approximately a factor of 100 from the simulated values, although the simulation parameters were chosen to match the expected material properties. In the simulations, we obtain not only the particle positions and center-of-mass velocities but also the angular velocities. From these, we determined the rotational MSDs (rMSDs), as described above. The rotational MSD differs from the translational MSD by approximately four orders of magnitude. The order of magnitude of the rMSD is comparable to the experimentally measured MSD. This suggests that the method used by Kunzner et al. \cite{kunzner2025dynamics} probes predominantly rotational rather than translational motion, as previously speculated.

\begin{figure}
    \centering
    \includegraphics[width=\linewidth]{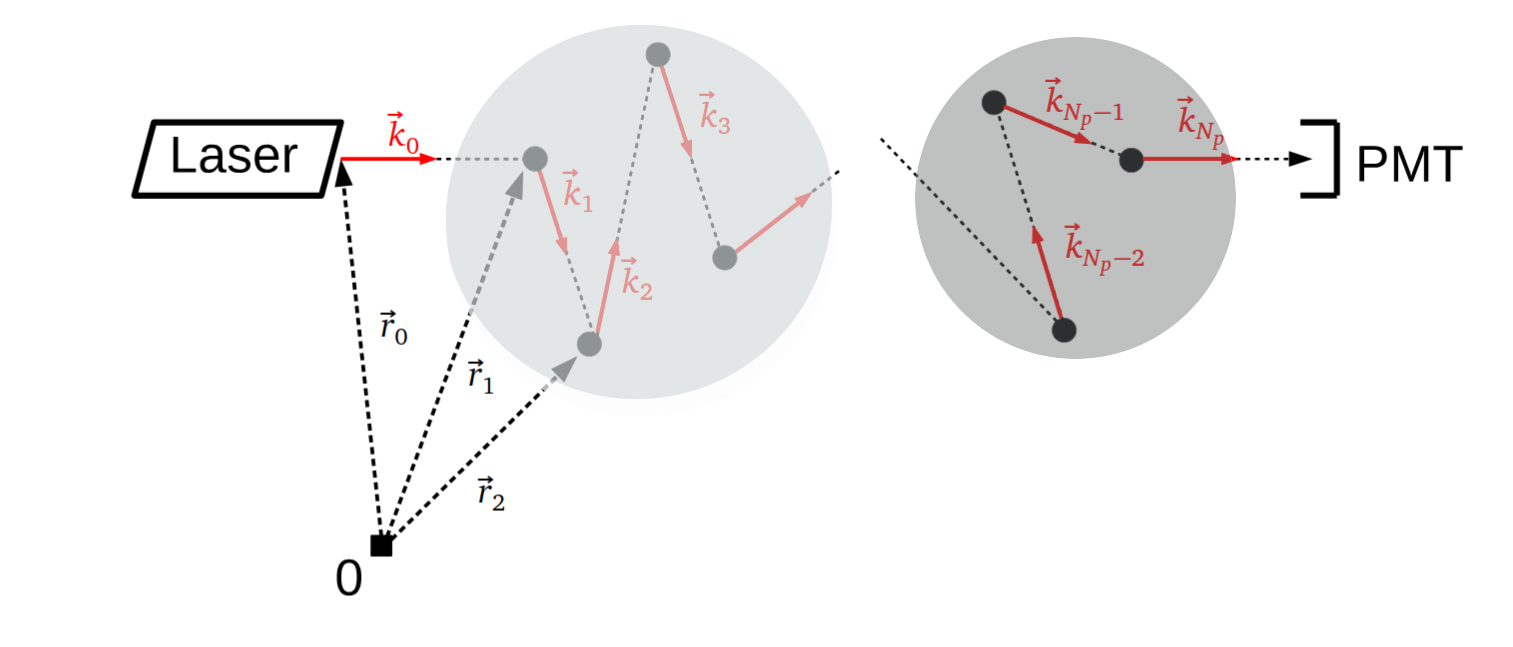}
    \caption{Schematic representation of the light path through a granular medium. The basis of the image is taken from \cite{M12_old}. As in this reference, $N_p$ is here the number of scattering centers, and PMT is the photomultiplier tube.}
    \label{fig:schema}
\end{figure}

To illustrate this we follow the derivation of the DWS technique in \cite{M12}, the electric field correlation function $g_1(t)$ can be obtained by superposing the contributions from all possible light propagation paths $p$, as illustrated in Fig.~\ref{fig:schema}:
\begin{equation}
 g_1(t) = \sum_p \frac{\langle E_p(0)\, E_p^\ast(t)\rangle}{\langle I \rangle} =\sum_p \frac{\langle I_p \rangle}{\langle I \rangle}\langle \ensuremath\mathrm{e}^{-i\,\Delta\Phi_p(t)} \rangle 
 \label{eq:feld2}
\end{equation}
where $E_p(t)$ denotes the electric field and $I_p(t)$ the intensity associated with an individual scattering path $p$. The superposition of all scattered spherical waves gives rise to the accumulated phase shift $\Delta\Phi_p(t)$ along path $p$ (schematically shown in Fig.~\ref{fig:schema}). Assuming elastic scattering, all wave vectors have the same magnitude, $k_i = 2\pi/\lambda$, while their directions vary from one scattering event to the next:
\begin{equation}
\Phi_p(t)=\sum_{i=0}^{N_p} \vec{k}_i(t)[\vec{r}_{i+1}(t)-\vec{r}_{i}(t)]
\label{eq:phi}
\end{equation}
For conventional DWS, where each scattering event originates from an independent particle, the accumulated phase fluctuations are directly related to the translational mean-squared displacement (MSD). In the present system, however, the scattering centers are distributed within the volume and on the surface of the particles. A rigid-body translation shifts all scattering centers by the same displacement, leaving their relative positions unchanged. Consequently, scattering events occurring within the same particle do not contribute to the phase decorrelation associated with translational motion. In contrast, a particle rotation changes the relative positions of all internal scattering centers with respect to the not rotated scattered wave vectors, leading to a finite phase shift and, therefore, to a rotational contribution to the measured MSD. Only scattering paths that connect different particles are sensitive to the relative translational motion between particles. Since multiple scattering  contains a combination of intra- and inter-particle scattering events, the measured MSD represents a superposition of translational and rotational dynamics with an unknown weighting.

\section{Discussion}

The DWS technique relates the intensity autocorrelation of multiply scattered light to the mean-squared displacement of the scattering centers \cite{Brown}. Following the derivation of the DWS approximation \cite{Brown}, it is plausible that the measured quantity for polystyrene particles is a combination of translational and rotational contributions (MSD and rMSD). Since the Dynoseed particles used in the experiment are white, they contain numerous air bubbles, which act as additional scattering centers inside the particles and likely dominate the DWS signal. Therefore, the relative positions of the air bubbles within each particle are primarily probed. The measured MSD thus reflects not only interparticle translation but also particle rotation, which repositions the air bubbles within the scattering volume (Dynoseed particle). Additionally, in experiments, we expect a significant degree of polydispersity. Since the rMSD contains the particle diameter as a scaling factor, variations in particle size are expected to affect the magnitude of the rotational contribution and therefore influence the measured signal.

\begin{figure}[t!]
\centering
\includegraphics[width=0.95\linewidth]{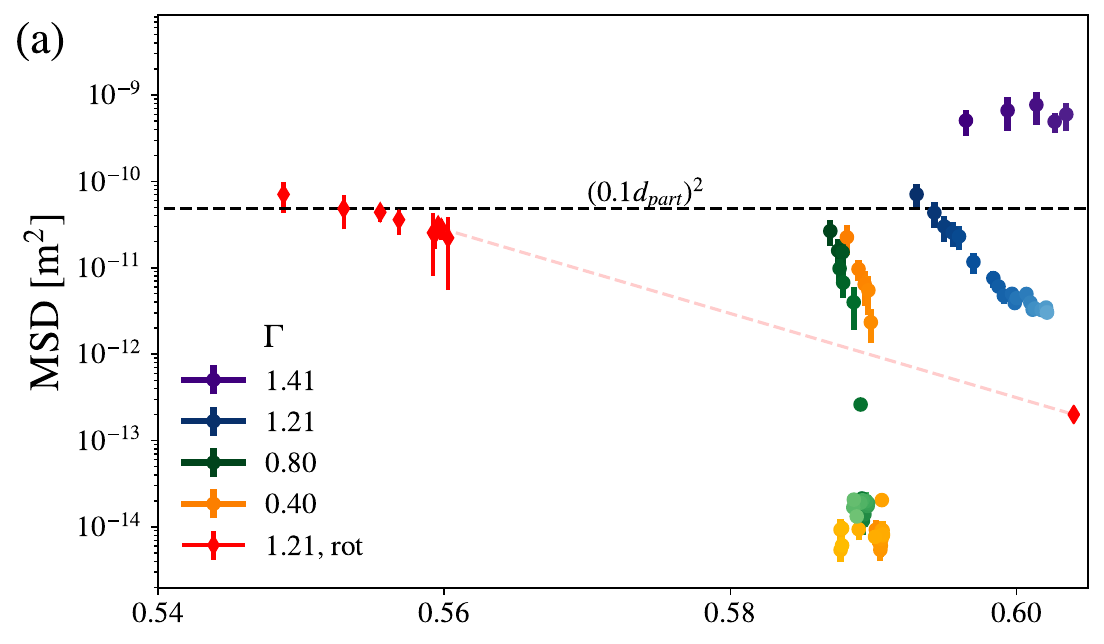}
\includegraphics[width=0.95\linewidth]{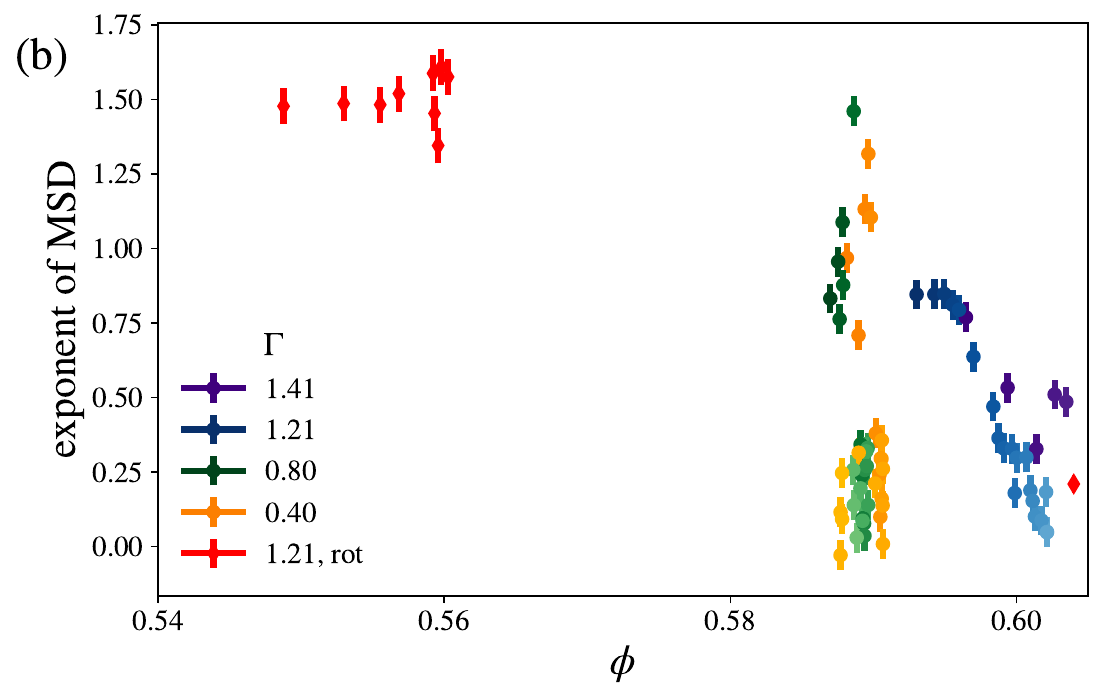}
\caption{Density-dependent parameter deduced from MSD at 0.1~s for agitation amplitudes between $\Gamma = 1.4$ and $\Gamma = 0.4$ without rotational friction (circles) and with rotational friction (diamonds) (a) MSDs and (b) power-law exponents. The dashed line indicates the Lindemann length. The error bars represent the standard errors of the fitted parameters obtained from the corresponding fits. }
\label{fig:parameter}
\end{figure}

To quantify the time evolution of the dynamics during densification, we analyze the MSD values at approximately 0.1~s and the power-law exponent, fitted at $t>5 \times 10^{-1}$, following the onset of caging (Fig.~\ref{fig:parameter}). The MSD value reflects, on the one hand, the available free volume and, on the other hand, the magnitude of the particle dynamics. 

The densification in reality takes approximately 100 times longer than in simulations without rotational friction. The time required to densify in simulations with friction is of the same order as in the experiment. Surprisingly, the simulation with friction stops densifying earlier despite continued shaking. Possible influences include the missing implementation of polydispersity \cite{herrmann1998modeling,an2011study}, the fact that the simulated system is only half the size of the experimental system, and the simplicity of the interaction model, which may be insufficient to reproduce the full time dependence of the densification process. The real system does not follow simple spring interactions, may exhibit charging effects, and likely possesses a much more complex energy landscape governing the time-dependent densification.

The superdiffusive behavior of the rMSD is, at first glance, surprising. It remains superdiffusive while the translational MSD evolves from subdiffusive to constant during the slowdown introduced by the densification. Superdiffusive behavior has, for example, been observed in rheological experiments by Scalliet et al. \cite{Scalliet2015Cages} and was associated with creep. Vibrated fluidized beds are often discussed in the context of convection rolls \cite{schroter2006mechanisms, Stephan2007Influence, Keiko1996Convective, Garcimart2002Convective, MAJID2009311}. However, it is not immediately clear how convection could explain the observed behavior, since it is expected to influence the translational MSD as well. 

\begin{figure}[t!]
    \centering
    \includegraphics[width=0.9\linewidth]{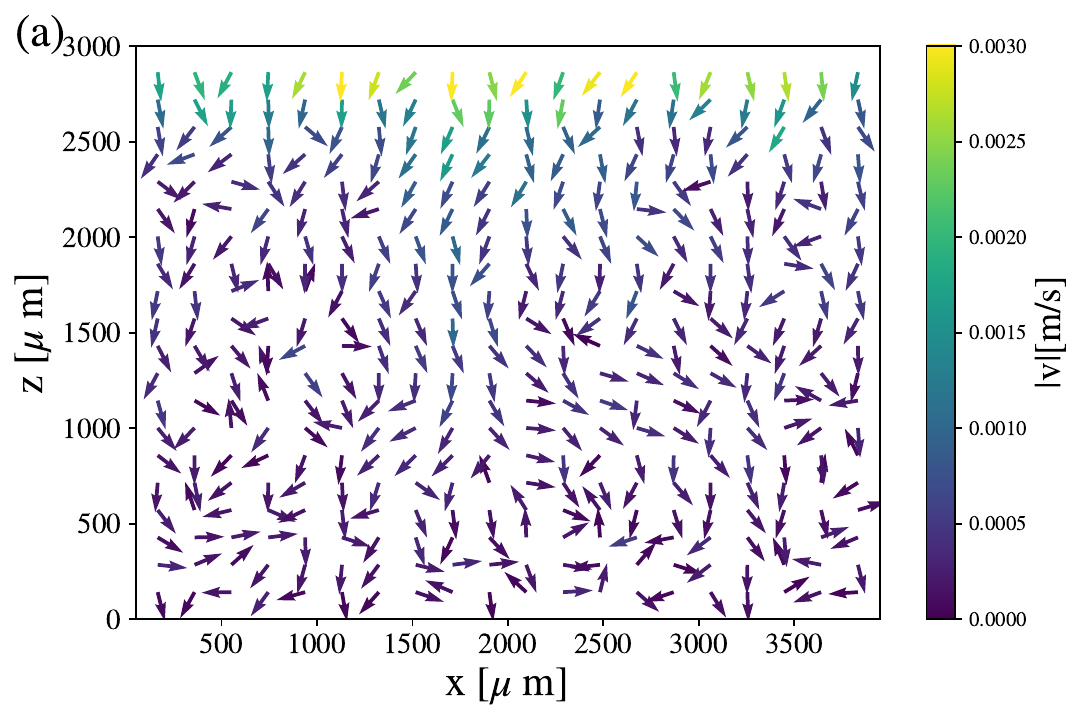}
    \includegraphics[width=0.9\linewidth]{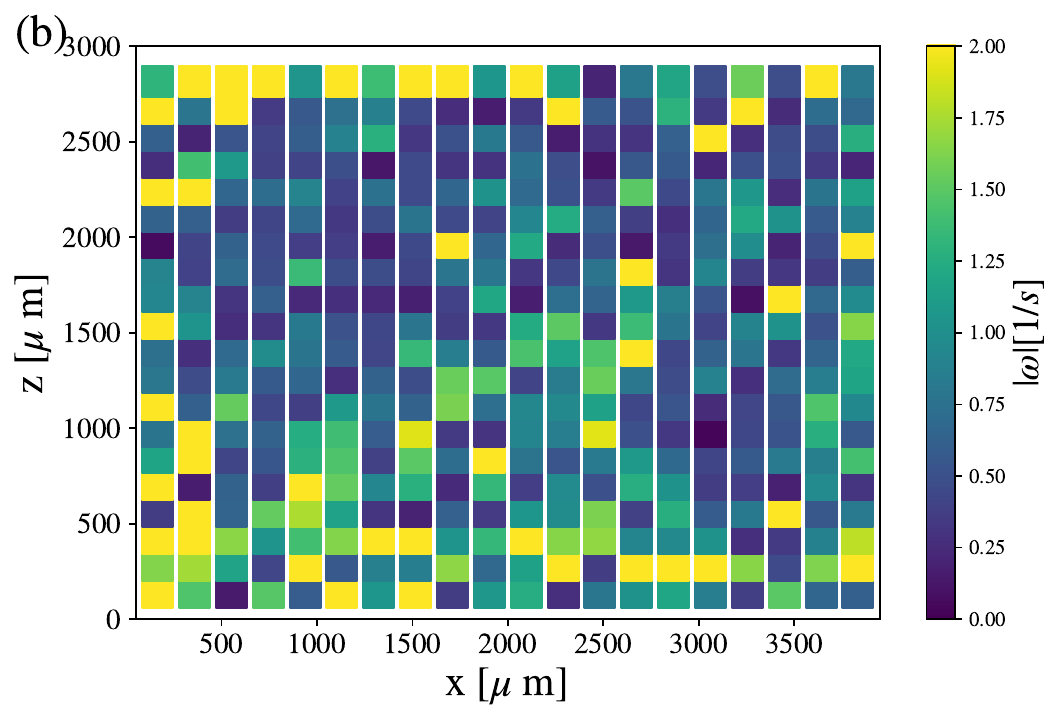}
    \includegraphics[width=0.9\linewidth]{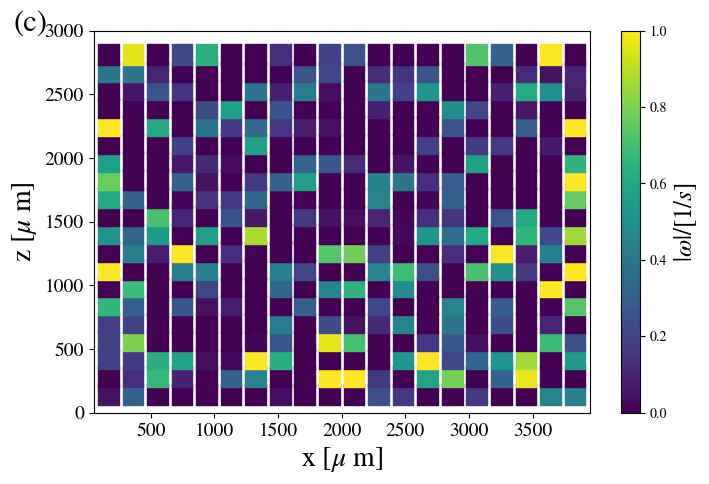}
    \caption{Illustration of the velocity fields in the simulations, averaged over 200 timesteps : (a) shows the translational velocity field, from which the affine motion is subtracted. (b) shows the velocity field of the \emph{angular} velocity in the $xz$-plane. (c) shows the $\omega_y$-component.  
    }
    \label{fig:velocetyfield}
\end{figure}

For further analysis, we show in Fig.~\ref{fig:velocetyfield} the translational and rotational velocity fields during the last 200~s of the densification process for the system with rotational friction. While the translational field indicates overall densification and minor rotational motion near the bottom, directed motion is observed along the walls and in the center of the rotational field. We therefore conclude that densification in the dense state proceeds without pronounced convection rolls, but that the walls induce directed rotational motion, leading to superdiffusive rMSD behavior. This behavior eventually ceases as the system approaches its highest packing fractions and evolves into diffusive rMSD and subdiffusive MSD.

It is important to note that $k_t$	and $\gamma_t$ dictate the exact scale and slope of the rotational response. Because these tangential contact parameters are difficult to measure directly for microscopic polystyrene spheres, the present DEM results should be viewed as a physical mechanism for the DWS signal rather than a fitted parameter study. Conversely, matching simulated rMSDs against experimental DWS curves offers a novel path to inversely calibrate tangential contact parameters in dense, vibrated granular media.

\section{Summary}

In this paper, we investigated the dynamics of granular matter during vibration-induced densification using DEM simulations. In agreement with experiments and theoretical expectations, we found that the packing fraction of the system increases during the shaking process. This increase depends on the vibration amplitude, and the final packing fraction varies with amplitude.

We further analyzed the system dynamics during densification by studying the MSD. The simulations reproduce the qualitative dynamical behavior observed in the experiments. The origin of the discrepancy between simulated and experimentally measured MSDs is discussed, and we find that the order of magnitude of the rotational MSD (rMSD) is comparable to the MSD measured experimentally. 
This suggests that in systems with internal particle scattering centers, experimental DWS signals are strongly sensitive to rotational motion, offering a plausible physical mechanism for the ultra-small caging lengths observed near glassification. These findings reinterpret previously reported ultra-small caging lengths and provide a consistent physical explanation for DWS observations in vibrated granular systems.

In future work, the rotational MSD will be investigated in more detail. In particular, the influence of gravitational strength and wall interactions will be studied. The densification in more complex agitation scenarios in microgravity can also be addressed, including the interpretation of light-scattering experiments such as those reported by Mayo et al. \cite{mayo2025observing}, where the measured signal may reflect particle reorientation.
The role of the tangential stiffness and damping parameter on the magnitude of the rMSD will be studied and Experiments with coated spheres reducing the friction are already in preparation.

Furthermore, the creep conditions under which superdiffusive rotational motion occurs remain incompletely understood and require systematic investigation. Such studies would improve the understanding of the microscopic mechanisms governing granular dynamics during densification.

Finally, a more detailed calibration of the simulation parameters is necessary to achieve a quantitative description of the rotational dynamics and the time-dependent densification behavior of the polystyrene system.

\section{Data availability}
The data that support the findings of this article are openly available \cite{zenodo}.

\section{Acknowledgments}
The authors acknowledge fruitful discussions with Timo Plath, Marlo Kunzner, Matthias Sperl, and Thomas Voigtmann. The authors gratefully acknowledge the scientific support and HPC resources provided by the German Aerospace Center (DLR). The HPC system CARO is partially funded by ``Ministry of Science and Culture of Lower Saxony'' and ``Federal Ministry for Economic Affairs and Climate Action''. We are grateful for the support of ESA and the German Federal Parliament under grant number 50WM1945 (SoMaDy2).

\section{Funding}
The work presented was supported by German Federal Parliament under grant number 50WM1945 (SoMaDy2)

\section{Contributions}
L.H. and J.P.G. wrote the main manuscript and analysed the data. L.H. prepared the figures and performed the simulations. J.P.G. supervised the work. All authors reviewed the manuscript.

\section{Conflict of interest}
The authors declare that they have no conflict of interest.
\bibliography{DEM_VC}

@article{d2021gravity,
  title={A gravity-independent powder-based additive manufacturing process tailored for space applications},
  author={D’Angelo, Olfa and Kuthe, Felix and Liu, Szu-Jia and Wiedey, Raphael and Bennett, Joe M and Meisnar, Martina and Barnes, Andrew and Kranz, W Till and Voigtmann, Thomas and Meyer, Andreas},
  journal={Addit. Manuf.},
  volume={47},
  pages={102349},
  year={2021},
  publisher={Elsevier}
}

@article{mair2007nature,
  title={{Nature of stress accommodation in sheared granular material: Insights from 3D numerical modeling}},
  author={Mair, Karen and Hazzard, James F},
  journal={Earth Planet. Sci. Lett.},
  volume={259},
  number={3-4},
  pages={469--485},
  year={2007},
  publisher={Elsevier}
}

@article{vescovi2016merging,
  title={Merging fluid and solid granular behavior},
  author={Vescovi, Dalila and Luding, Stefan},
  journal={Soft matter},
  volume={12},
  number={41},
  pages={8616--8628},
  year={2016},
  publisher={Royal Society of Chemistry}
}

@article{malone2008determination,
  title={Determination of contact parameters for discrete element method simulations of granular systems},
  author={Malone, Kevin Francis and Xu, Bao Hua},
  journal={Particuology},
  volume={6},
  number={6},
  pages={521--528},
  year={2008},
  publisher={Elsevier}
}

@article{brilliantov1996model,
  title={Model for collisions in granular gases},
  author={Brilliantov, Nikolai V and Spahn, Frank and Hertzsch, Jan-Martin and P{\"o}schel, Thorsten},
  journal={Phys. Rev. E},
  volume={53},
  number={5},
  pages={5382},
  year={1996},
  publisher={APS}
}

@article{silbert2001granular,
  title={Granular flow down an inclined plane: Bagnold scaling and rheology},
  author={Silbert, Leonardo E and Erta{\c{s}}, Deniz and Grest, Gary S and Halsey, Thomas C and Levine, Dov and Plimpton, Steven J},
  journal={Phys. Rev E},
  volume={64},
  number={5},
  pages={051302},
  year={2001},
  publisher={APS}
}

@article{zhang2005jamming,
  title={Jamming transition in emulsions and granular materials},
  author={Zhang, HP and Makse, HA},
  journal={Phys. Rev. E.},
  volume={72},
  number={1},
  pages={011301},
  year={2005},
  publisher={APS}
}

@article{elizondo2020arrested,
  title={Arrested dynamics of the dipolar hard sphere model},
  author={Elizondo-Aguilera, Luis F and Cort{\'e}s-Morales, Ernesto C and Rico, Pablo F Zubieta and Medina-Noyola, Magdaleno and Casta{\~n}eda-Priego, Ram{\'o}n and Voigtmann, Thomas and P{\'e}rez-{\'A}ngel, Gabriel},
  journal={Soft matter},
  volume={16},
  number={1},
  pages={170--190},
  year={2020},
  publisher={Royal Society of Chemistry}
}

@article{di2004comparison,
  title={Comparison of contact-force models for the simulation of collisions in DEM-based granular flow codes},
  author={Di Renzo, Alberto and Di Maio, Francesco Paolo},
  journal={Chem. Eng. Sci.},
  volume={59},
  number={3},
  pages={525--541},
  year={2004},
  publisher={Elsevier}
}

@article{kunzner2025dynamics,
  title={Dynamics in vibrofluidized beds: A diffusing wave spectroscopy study},
  author={Kunzner, Marlo and Mayo, Christopher and Sperl, Matthias and Gabriel, Jan Philipp},
  journal={Phys. Rev. E},
  volume={112},
  number={2},
  pages={025405},
  year={2025},
  publisher={APS}
}

@article{jaeger1996physics,
  title={The physics of granular materials},
  author={Jaeger, Heinrich M and Nagel, Sidney R and Behringer, Robert P},
  journal={Phys. today},
  volume={49},
  number={4},
  pages={32--38},
  year={1996},
  publisher={American Institute of Physics}
}

@book{duran2012sands,
  title={Sands, powders, and grains: an introduction to the physics of granular materials},
  author={Duran, Jacques},
  year={2012},
  publisher={Springer Science \& Business Media}
}

@article{dufresne2012granular,
  title={Granular flow experiments on the interaction with stationary runout path materials and comparison to rock avalanche events},
  author={Dufresne, Anja},
  journal={Earth Surf. Process. Landforms},
  volume={37},
  number={14},
  pages={1527--1541},
  year={2012},
  publisher={Wiley Online Library}
}

@article{rosato1987brazil,
  title={Why the Brazil nuts are on top: Size segregation of particulate matter by shaking},
  author={Rosato, Anthony and Strandburg, Katherine J and Prinz, Friedrich and Swendsen, Robert H},
  journal={Phys. Rev. Lett.},
  volume={58},
  number={10},
  pages={1038},
  year={1987},
  publisher={APS}
}

@article{siemens2010jamming,
  title={Jamming: A simple introduction},
  author={Siemens, Alexander ON and Van Hecke, Martin},
  journal={Physica A},
  volume={389},
  number={20},
  pages={4255--4264},
  year={2010},
  publisher={Elsevier}
}

@article{behringer2018physics,
  title={The physics of jamming for granular materials: a review},
  author={Behringer, Robert P and Chakraborty, Bulbul},
  journal={Rep. Prog. Phys.},
  volume={82},
  number={1},
  pages={012601},
  year={2018},
  publisher={IOP Publishing}
}

@article{zhang2010jamming,
  title={Jamming for a {2D} granular material},
  author={Zhang, Jie and Majmudar, TS and Sperl, Matthias and Behringer, RP},
  journal={Soft Matter},
  volume={6},
  number={13},
  pages={2982--2991},
  year={2010},
  publisher={Royal Society of Chemistry}
}

@article{kudrolli2004size,
  title={Size separation in vibrated granular matter},
  author={Kudrolli, Arshad},
  journal={Rep. Prog. Phys.},
  volume={67},
  number={3},
  pages={209},
  year={2004},
  publisher={IOP Publishing}
}

@article{kollmer2020migrating,
  title={Migrating shear bands in shaken granular matter},
  author={Kollmer, Jonathan E and Shreve, Tara and Claussen, Joelle and Gerth, Stefan and Salamon, Michael and Uhlmann, Norman and Schr{\"o}ter, Matthias and P{\"o}schel, Thorsten},
  journal={Phys. Rev. Lett.},
  volume={125},
  number={4},
  pages={048001},
  year={2020},
  publisher={APS}
}

@article{eshuis2007phase,
    author = {Eshuis, Peter and van der Weele, Ko and van der Meer, Devaraj and Bos, Robert and Lohse, Detlef},
    title = {Phase diagram of vertically shaken granular matter},
    journal = {Physics of Fluids},
    volume = {19},
    number = {12},
    pages = {123301},
    year = {2007},
    month = {12},
    issn = {1070-6631},
    doi = {10.1063/1.2815745},
    url = {https://doi.org/10.1063/1.2815745},
}

@article{el2021theories,
  title={{Theories and applications of CFD--DEM coupling approach for granular flow: A review}},
  author={El-Emam, Mahmoud A and Zhou, Ling and Shi, Weidong and Han, Chen and Bai, Ling and Agarwal, Ramesh},
  journal={Arch. Comput. Methods Eng.},
  volume={28},
  number={7},
  pages={4979--5020},
  year={2021},
  publisher={Springer}
}

@article{hadi2024modelling,
  title={{DEM} modelling of segregation in granular materials: a review},
  author={Hadi, Ahmed and Roeplal, Ra{\"\i}sa and Pang, Yusong and Schott, Dingena L},
  journal={KONA Powder Part. J.},
  volume={41},
  pages={78--107},
  year={2024},
  publisher={Hosokawa Powder Technology Foundation}
}

@article{di2021coarse,
  title={Coarse-grain dem modelling in fluidized bed simulation: A review},
  author={Di Renzo, Alberto and Napolitano, Erasmo S and Di Maio, Francesco P},
  journal={Processes},
  volume={9},
  number={2},
  pages={279},
  year={2021},
  publisher={MDPI}
}

@Article{LAMMPS,
  author = "A. P. Thompson and H. M. Aktulga and R. Berger and 
     D. S. Bolintineanu and W. M. Brown and P. S. Crozier and
     P. J. in 't Veld and A. Kohlmeyer and S. G. Moore and T. D. Nguyen and
     R. Shan and M. J. Stevens and J. Tranchida and C. Trott and S. J. Plimpton",
  title = "{LAMMPS} - a flexible simulation tool for
     particle-based materials modeling at the 
     atomic, meso, and continuum scales",
  journal = "Comp. Phys. Comm.",
  volume =  "271",
  pages =   "108171",
  year =    "2022",
  doi = "10.1016/j.cpc.2021.108171"
}

@article{sweeney2017characterizing,
  title={Characterizing the feasibility of processing wet granular materials to improve rheology for {3D} printing},
  author={Sweeney, Michael and Campbell, Loudon L and Hanson, Jeff and Pantoya, Michelle L and Christopher, Gordon F},
  journal={J. Mater. Sci.},
  volume={52},
  number={22},
  pages={13040--13053},
  year={2017},
  publisher={Springer}
}

@article{roy2024role,
  title={The role of granular matter in additive manufacturing},
  author={Roy, Sudeshna and Weinhart, Thomas},
  journal={Granul. Matter},
  volume={26},
  number={4},
  pages={102},
  year={2024},
  publisher={Springer}
}

@article{asaf2024granular,
  title={Granular materials for {3D} printing of construction components and structures},
  author={Asaf, Ofer and Bentur, Arnon and Larianovsky, Pavel and Sprecher, Aaron},
  journal={Autom. Constr.},
  volume={166},
  pages={105544},
  year={2024},
  publisher={Elsevier}
}

@manual{lammpsmsd,
    title = {Calculation of the Mean-Squared Displacement},
    note= {\url{https://docs.lammps.org/compute_msd.html}}
}

@article{Scalliet2015Cages,
  title = {Cages and Anomalous Diffusion in Vibrated Dense Granular Media},
  author = {Scalliet, Camille and Gnoli, Andrea and Puglisi, Andrea and Vulpiani, Angelo},
  journal = {Phys. Rev. Lett.},
  volume = {114},
  issue = {19},
  pages = {198001},
  numpages = {5},
  year = {2015},
  month = {May},
  publisher = {American Physical Society},
  doi = {10.1103/PhysRevLett.114.198001},
  url = {https://link.aps.org/doi/10.1103/PhysRevLett.114.198001}
}

@book{Brown,
	author = {W. Brown},
	title = {Dynamic Light Scattering},
	date = {1993},
	editor = {W. Brown},
	publisher = {Clarendon Press},
	chapter = {16},
	addendum = {Diffusing Wave Spektroskopie},
}

@article{schroter2006mechanisms,
  title={Mechanisms in the size segregation of a binary granular mixture},
  author={Schr{\"o}ter, Matthias and Ulrich, Stephan and Kreft, Jennifer and Swift, Jack B and Swinney, Harry L},
  journal={Physical Review E—Statistical, Nonlinear, and Soft Matter Physics},
  volume={74},
  number={1},
  pages={011307},
  year={2006},
  publisher={APS}
}

@article{Stephan2007Influence,
  title = {Influence of friction on granular segregation},
  author = {Ulrich, Stephan and Schr\"oter, Matthias and Swinney, Harry L.},
  journal = {Phys. Rev. E},
  volume = {76},
  issue = {4},
  pages = {042301},
  numpages = {3},
  year = {2007},
  month = {Oct},
  publisher = {American Physical Society},
  doi = {10.1103/PhysRevE.76.042301},
  url = {https://link.aps.org/doi/10.1103/PhysRevE.76.042301}
}

@article{Keiko1996Convective,
  title = {Convective roll patterns in vertically vibrated beds of granules},
  author = {Aoki, Keiko M. and Akiyama, Tetsuo and Maki, Yoji and Watanabe, Tatsuyuki},
  journal = {Phys. Rev. E},
  volume = {54},
  issue = {1},
  pages = {874--883},
  numpages = {0},
  year = {1996},
  month = {Jul},
  publisher = {American Physical Society},
  doi = {10.1103/PhysRevE.54.874},
  url = {https://link.aps.org/doi/10.1103/PhysRevE.54.874}
}

@article{Garcimart2002Convective,
  title = {Convective motion in a vibrated granular layer},
  author = {Garcimart\'{\i}n, A. and Maza, D. and Ilquimiche, J. L. and Zuriguel, I.},
  journal = {Phys. Rev. E},
  volume = {65},
  issue = {3},
  pages = {031303},
  numpages = {5},
  year = {2002},
  month = {Feb},
  publisher = {American Physical Society},
  doi = {10.1103/PhysRevE.65.031303},
  url = {https://link.aps.org/doi/10.1103/PhysRevE.65.031303}
}

@article{MAJID2009311,
title = {Convection and segregation in vertically vibrated granular beds},
journal = {Powder Technol.},
volume = {192},
number = {3},
pages = {311-317},
year = {2009},
issn = {0032-5910},
doi = {https://doi.org/10.1016/j.powtec.2009.01.012},
url = {https://www.sciencedirect.com/science/article/pii/S003259100900059X},
author = {M. Majid and P. Walzel},
}

@misc{zenodo,
  author = {Heitmeier, Linnea and Gabriel, Jan},
  title = {Simulation Data},
  doi = {10.5281/zenodo.18681170},
  publisher = {Zenodo},
  year = {2026},
note={{DOI:}10.5281/zenodo.18681170},
}

@article{sperl2005nearly,
  title={Nearly logarithmic decay in the colloidal hard-sphere system},
  author={Sperl, Matthias},
  journal={Phys. Rev. E},
  volume={71},
  number={6},
  pages={060401},
  year={2005},
  publisher={APS}
}

@article{sperl2012single,
  title={Single-particle dynamics in dense granular fluids under driving},
  author={Sperl, Matthias and Kranz, W Till and Zippelius, Annette},
  journal={EPL},
  volume={98},
  number={2},
  pages={28001},
  year={2012},
  publisher={IOP Publishing}
}

@article{jerkins2008onset,
  title={Onset of mechanical stability in random packings of frictional spheres},
  author={Jerkins, Melissa and Schr{\"o}ter, Matthias and Swinney, Harry L and Senden, Tim J and Saadatfar, Mohammad and Aste, Tomaso},
  journal={Phys. Rev. Lett.},
  volume={101},
  number={1},
  pages={018301},
  year={2008},
  publisher={APS}
}

@article{liu2017equation,
  title={Equation of state for random sphere packings with arbitrary adhesion and friction},
  author={Liu, Wenwei and Jin, Yuliang and Chen, Sheng and Makse, Hern{\'a}n A and Li, Shuiqing},
  journal={Soft Matter},
  volume={13},
  number={2},
  pages={421--427},
  year={2017},
  publisher={Royal Society of Chemistry}
}

@article{mayo2025observing,
  title={Observing the glass and jamming transitions of dense granular material in microgravity},
  author={Mayo, Christopher and Kunzner, Marlo and Sperl, Matthias and Gabriel, Jan Philipp},
  journal={Phys. Rev. Lett.},
  volume={135},
  number={17},
  pages={178203},
  year={2025},
  publisher={APS}
}

@article{hashemnia2018study,
  title={Study the effect of vibration frequency and amplitude on the quality of fluidization of a vibrated granular flow using discrete element method},
  author={Hashemnia, K and Pourandi, S},
  journal={Powder Technol.},
  volume={327},
  pages={335--345},
  year={2018},
  publisher={Elsevier}
}

@book{berne2000dynamic,
  title={Dynamic light scattering: with applications to chemistry, biology, and physics},
  author={Berne, Bruce J and Pecora, Robert},
  year={2000},
  publisher={Courier Corporation}
}

@MISC{M12_old, 
    author       = {Blochowicz, Thomas}, 
    title        = {Jamming in granularen {M}edien}, 
    howpublished = {Private Communication}, 
    note         = {{U}niversity of {D}armstadt},
    year         = {2015},
    publisher    = {{Grundpraktikum TU Darmstadt}},
}

@article{liu1998jamming,
  title={Jamming is not just cool any more},
  author={Liu, Andrea J and Nagel, Sidney R},
  journal={Nature},
  volume={396},
  number={6706},
  pages={21--22},
  year={1998},
  publisher={Nature Publishing Group UK London}
}

@Inbook{berthier2009glasses,
author="Berthier, Ludovic
and Biroli, Giulio",
editor="Meyers, Robert A.",
title="Glasses and Aging, A Statistical Mechanics Perspective on",
bookTitle="Encyclopedia of Complexity and Systems Science",
year="2009",
publisher="Springer New York",
address="New York, NY",
pages="4209--4240",
isbn="978-0-387-30440-3",
doi="10.1007/978-0-387-30440-3_248",
url="https://doi.org/10.1007/978-0-387-30440-3_248"
}

@article{zivkovic2011scaling,
  title={Scaling of granular temperature in a vibrated granular bed},
  author={Zivkovic, V and Biggs, MJ and Glass, DH},
  journal={Phys. Rev. E},
  volume={83},
  number={3},
  pages={031308},
  year={2011},
  publisher={APS}
}

@article{kim2005jamming,
  title={Jamming transition in a highly dense granular system under vertical vibration},
  author={Kim, Kipom and Moon, Jong Kyun and Park, Jong Jin and Kim, Hyung Kook and Pak, Hyuk Kyu},
  journal={Phys. Rev. E},
  volume={72},
  number={1},
  pages={011302},
  year={2005},
  publisher={APS}
}

@article{kim2002solid,
  title={Solid-liquid transition in a highly dense 3D vibro-fluidized granular system},
  author={Kim, Kipom and Park, Jong Jin and Moon, Jong Kyun and Kim, Hyung Kook and Pak, Hyuk Kyu},
  journal={J. Korean Phys. Soc.},
  volume={40},
  pages={983--986},
  year={2002},
  publisher={Korean Physical Society; 1999}
}

@article{hanotin2013dynamics,
  title={Dynamics of vibrated granular suspensions probed by mechanical spectroscopy and diffusing wave spectroscopy measurements},
  author={Hanotin, Caroline and Marchal, Philippe and Michot, Laurent J and Baravian, Christophe and de Richter, S{\'e}bastien Kiesgen},
  journal={Soft Matter},
  volume={9},
  number={39},
  pages={9352--9360},
  year={2013},
  publisher={Royal Society of Chemistry}
}

@article{biggs2008granular,
  title={Granular temperature in a gas fluidized bed},
  author={Biggs, Mark J and Glass, Don and Xie, Liansong and Zivkovic, Vladimir and Buts, Alex and Curt Kounders, MA},
  journal={Granul. Matter},
  volume={10},
  number={2},
  pages={63--73},
  year={2008},
  publisher={Springer}
}

@article{menon1997particle,
  title={Particle motions in a gas-fluidized bed of sand},
  author={Menon, Narayanan and Durian, Douglas J},
  journal={Phys. Rev.},
  volume={79},
  number={18},
  pages={3407},
  year={1997},
  publisher={APS}
}

@article{menon1997diffusing,
  title={Diffusing-wave spectroscopy of dynamics in a three-dimensional granular flow},
  author={Menon, Narayanan and Durian, Douglas J},
  journal={Science},
  volume={275},
  number={5308},
  pages={1920--1922},
  year={1997},
  publisher={American Association for the Advancement of Science}
}

@article{herrmann1998modeling,
  title={Modeling granular media on the computer},
  author={Herrmann, HJ and Luding, Stefan},
  journal={Continuum Mechanics and Thermodynamics},
  volume={10},
  number={4},
  pages={189--231},
  year={1998},
  publisher={Springer}
}

@article{an2011study,
  title={DEM study of crystallization of monosized spheres under mechanical vibrations},
  author={An, Xizhong and Yang, Runyu and Dong, Kejun and Yu, Aibing},
  journal={Computer Physics Communications},
  volume={182},
  number={9},
  pages={1989--1994},
  year={2011},
  publisher={Elsevier}
}

@article{hanotin2015viscoelasticity,
  title={Viscoelasticity of vibrated granular suspensions},
  author={Hanotin, C and Kiesgen de Richter, S and Michot, LJ and Marchal, Ph},
  journal={J. Rheo.},
  volume={59},
  number={1},
  pages={253--273},
  year={2015},
  publisher={AIP Publishing}
}

@article{Lindemann1910,  author  = {Lindemann, F. A.},  title   = {The Calculation of Molecular Vibration Frequencies},  journal = {Physikalische Zeitschrift},  year    = {1910},  volume  = {11},  pages   = {609},}

@article{nicolas2000compaction,
  title={Compaction of a granular material under cyclic shear},
  author={Nicolas, Maxime and Duru, P and Pouliquen, Olivier},
  journal={Eur. Phys. J. E},
  volume={3},
  number={4},
  pages={309--314},
  year={2000},
  publisher={Springer}
}

@article{knight1995density,
  title={Density relaxation in a vibrated granular material},
  author={Knight, James B and Fandrich, Christopher G and Lau, Chun Ning and Jaeger, Heinrich M and Nagel, Sidney R},
  journal={Phys. Rev. E},
  volume={51},
  number={5},
  pages={3957},
  year={1995},
  publisher={APS}
}

@article{nowak1998density,
  title={Density fluctuations in vibrated granular materials},
  author={Nowak, Edmund R and Knight, James B and Ben-Naim, Eli and Jaeger, Heinrich M and Nagel, Sidney R},
  journal={Physical Review E},
  volume={57},
  number={2},
  pages={1971},
  year={1998},
  publisher={APS}
}

@article{makse2000packing,
  title={Packing of compressible granular materials},
  author={Makse, Hern{\'a}n A and Johnson, David L and Schwartz, Lawrence M},
  journal={Phys. Rev. Lett.},
  volume={84},
  number={18},
  pages={4160},
  year={2000},
  publisher={APS}
}

@article{M12,
  title={Jamming and Flow in Granular Matter: A Physics Lab Course Experiment},
  author={Blochowicz, Thomas and Ismajli, Emina and Gabriel, Jan Philipp},
  journal={arXiv preprint arXiv:2603.26857},
  year={2026}
}

@article{gerlach1927messung,
  title={{\"U}ber die Messung der rotatorischen Brownschen Bewegung mit Hilfe einer Drehwage},
  author={Gerlach, Werner and Lehrer, E},
  journal={Naturwissenschaften},
  volume={15},
  number={1},
  pages={15--15},
  year={1927},
  publisher={Springer}
}

\end{document}